\documentclass[aps,prd,preprintnumbers,nofootinbib,showpacs,onecolumn]{revtex4}%

\usepackage[dvips,dvipdfmx]{graphicx}
\usepackage{bm,latexsym,amsmath,amssymb,amsfonts,mathrsfs}
\usepackage{color}
\usepackage{subfigure}

\newcommand{\red}[1]{{\red{blue} #1}}

\begin{document}

\title{Vacuum state in de~Sitter spacetime with static charts}

\author{Atsushi Higuchi}
\affiliation{Department of Mathematics, University of York, Heslington, York YO10 5DD, United Kingdom}

\author{Kazuhiro Yamamoto}
\affiliation{Department of Physics, Graduate School of Science, Hiroshima University,
  Higashi-Hiroshima 739-8526, Japan}

\begin{abstract}
We study the free massive scalar field in de Sitter spacetime with static charts.  In particular,
we find positive-frequency modes for the Bunch-Davies vacuum state natural to the static charts
as superpositions of the well-known positive-frequency modes in the conformally-flat chart.
We discuss in detail how these modes are defined globally in the two static charts and the
region in their future.  The global structure of these solutions leads to the well-known
description of the Bunch-Davies vacuum state as an entangled state.
Our results are expected to be useful not only for studying the thermal properties in the
vacuum fluctuations in de~Sitter spacetime but also for understanding the nonlocal properties
of the vacuum state.
\end{abstract}
\maketitle

\section{Introduction}
Quantum entanglement plays an important role in quantum field theory in curved spacetime
in some cases. Thermal properties that appear in vacuum states in curved spacetimes
can be understood as a result of entanglement of states in causally disconnected
regions. The Unruh effect is an example, which predicts that a uniformly accelerating
observer sees the Minkowski vacuum as a thermally excited state with the Unruh
temperature $T_U=a/2\pi$, where $a$ is the acceleration of the observer
\cite{Fulling,Davies,Unruh,BisognanoWichmann,UnruhWald,Higuchi,ref-4,BiD,Lohiya}.
The entanglement in the vacuum state of a field is important not only
for the thermal nature\footnote{{See Ref.~~\cite{VerchBuchholz} 
for a recent critical discussion of the Unruh thermal bath.}}
seen by a uniformly accelerating observer, but also
for the quantum radiation which appears as the result of the Unruh effect
\cite{HIUY,ITUY,IOTYZ}.
The authors of Refs.~\cite{HIUY,ITUY,IOTYZ} demonstrated that the entanglement
structure between the states of the left and right Rindler wedges
for describing the Minkowski vacuum state is essential in understanding the
quantum radiation produced by a uniformly accelerating detector.

It is also well known that thermal properties appear in the quantum field theory
in de~Sitter spacetime \cite{GH,BD,HiguchiA,MTY,BCH}.
A detector at rest in de~Sitter spacetime which is coupled to the vacuum
fluctuations shows the thermal excitation with the Gibbons-Hawking temperature
$T_{GH}=H/2\pi$, where $H$ is the Hubble parameter
that characterizes de~Sitter spacetime~\cite{GH}.
This phenomenon can also be understood in terms of entanglement between
the states in the two causally disconnected static regions~\cite{BiD,TS94,Markkanen}.
The entanglement structure of the vacuum states in de~Sitter spacetime is
important for understanding not only the thermal properties but also the quantum
radiation produced by a uniformly accelerating detector in de~Sitter spacetime
\cite{Yamaguchi1}.
This might also be important for understanding the nonlocal properties and the
quantum entanglement entropy of quantum field theory in the vacuum states
in de~Sitter spacetime \cite{Pimentel,KM,Noumi,KSS1,KSS2,Kanno,KST},
which might provide us with some insight into relativistic quantum information
\cite{KukitaNambu,Rotondo,KukitaNambu2,MatsumuraNambu} and holographic gravity
dual theories.

\begin{figure*}[t]
    \vspace{-1cm}
  \begin{center}
        \includegraphics[width=15cm]{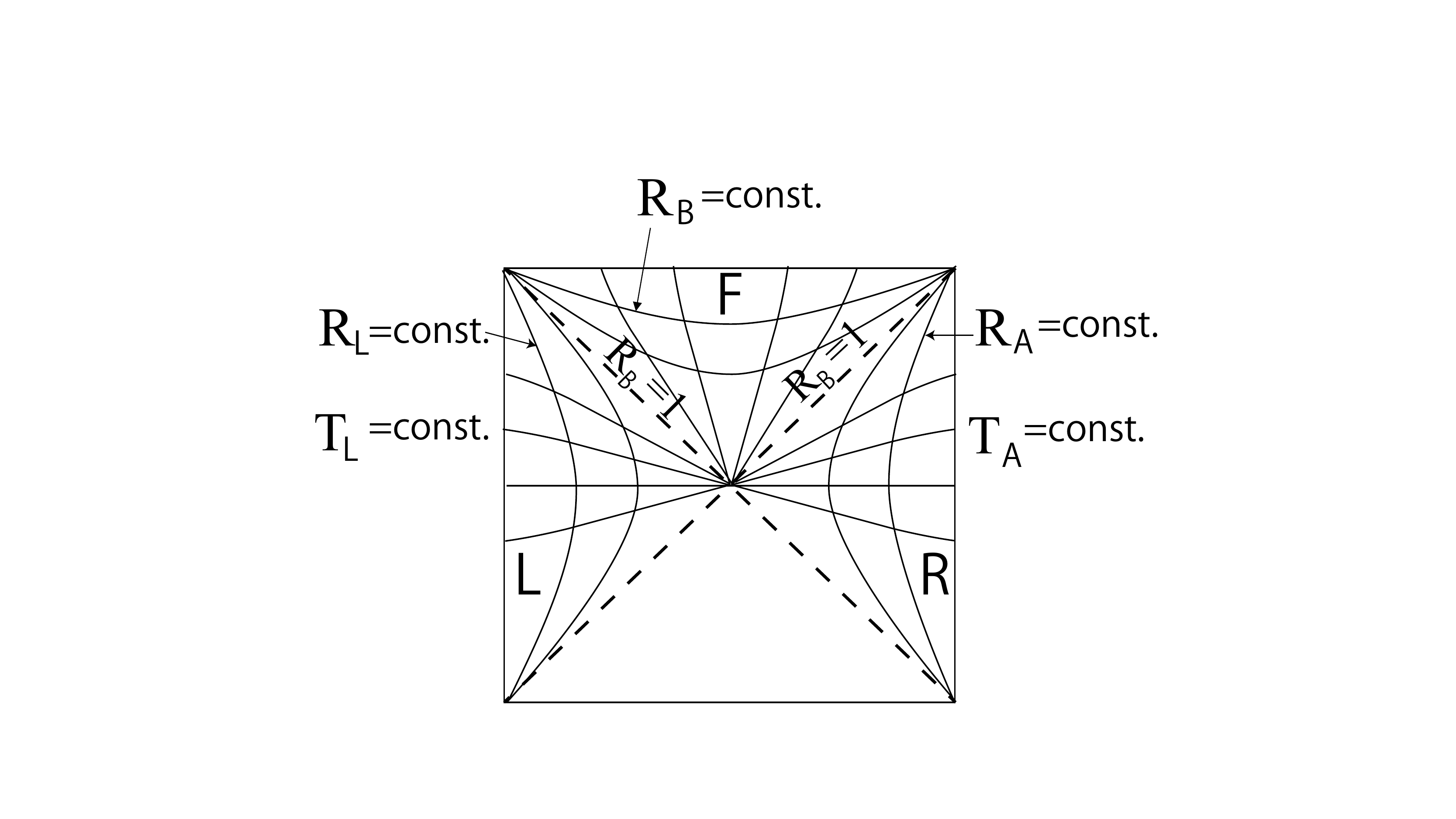}%
  \end{center}
  \vspace{-1cm}
\caption{
  \label{fig1} {
Conformal spacetime diagram of de~Sitter spacetime.
The R-region and the L-region are described by static chart
(see Secs.~III A and B), while the future F-region is describe
by an expanding universe (see Sec.~III C).}}
\end{figure*}

In this paper
we clarify the global properties of the positive-frequency modes, which lead
to this entanglement structure in the Bunch-Davies vacuum
state~\cite{GH,BD,ChernikovTagirov,ChomSpin}
in de~Sitter spacetime. In particular we construct positive-frequency modes suitable for
this purpose in the region to the future of the two static charts, which are then
analytically continued to the static charts. (See Fig.~\ref{fig1} for a
Carter-Penrose diagram of de~Sitter spacetime.)
These modes are analogous to the positive-frequency
modes in Minkowski spacetime natural to
the two Rindler wedges~\cite{Unruh,HIUY}.
Then we use the globally-defined positive-frequency modes thus obtained to understand
the entanglement structure of the vacuum state in de~Sitter spacetime.

The rest of the paper is organized as follows. In Sec.~\ref{sec:two} we summarize
the relations between the mode functions in Rindler wedges 
and those constructed in the future region described by the expanding
degenerate Kasner universe, which are connected by analytic continuation.
In Sec.~\ref{sec:three} we summarize various coordinate charts of de~Sitter spacetime
that we use in this paper.
In Sec.~\ref{sec:four} we construct positive-frequency modes
for the Bunch-Davies vacuum state in the future region as superpositions of the well-known
positive-frequency modes in the conformally-flat chart, also known as the Ponicar\'{e} chart
that contains the future region.
In Sec.~\ref{sec:five} we analytically continue these positive-frequency modes
to the left and right static charts,
and in Sec.~\ref{sec:six} we use this result to derive the expression for the
Bunch-Davies vacuum state as a state with entanglement between
states in the right static chart and those in the left static chart.
We also study analogous entanglement for the $\alpha$-vacua~\cite{BA,Mottola}
and find the entanglement entropy of a pair
of modes consisting of entangled states.
We find that the entanglement entropy does not depend on the mass of the field,
contrary to the previous works with two open charts~\cite{Pimentel,KM,Noumi}.

\begin{center}
\begin{table}[t]
\begin{tabular}{lll}
\hline\hline\\
~~~~~~
 ${\rm F} \longleftrightarrow {\rm R}$ ~~~&
 $\displaystyle{\zeta=\tau+{\pi \over 2 }i}$, ~~~~&
 $\displaystyle{\eta=\xi -{\pi \over 2}i}$~~~~~~
\\
\\
~~~~~~
 ${\rm F} \longleftrightarrow {\rm L}$ ~~~&
 $\displaystyle{\zeta = - \tilde \tau -{\pi \over 2 }i}$,~~~~&
 $\displaystyle{\eta = \tilde \xi - {\pi \over 2}i}$
\\
\\
\hline\hline
\end{tabular}
\label{tabletwo}
\caption{Analytic continuation for the coordinate variables between the R-region,
  the F-region and the L-region, in Minkowski spacetime.}
\end{table}
\end{center}

\section{Summary of entanglement structure in Minkowski spacetime} \label{sec:two}
We first review
the relationship between the mode functions for scalar field of mass $m$ constructed in Rindler spacetime and those in the
expanding degenerate Kasner spacetime as presented in Ref.~\cite{HIUY}. Then we present the well-known expression of the
Minkowski vacuum state as an entangled state, which is a consequence of this relationship.
This review will be useful for understanding the
relationship between the mode functions of the scalar quantum field in the static charts
and those in the future region of de~Sitter spacetime.
For simplicity we consider $2$-dimensional Minkowski spacetime with cartesian coordinates $t$ and $z$.  (In $4$-dimensions
the sector with transverse momentum $\bm{k}_\perp$ is equivalent to the $2$-dimensional counterpart with $m$ changed
to $\sqrt{m^2 + \bm{k}_\perp^2}$.)

We first summarize the coordinate systems we use for each region of spacetime:
\begin{itemize}
\item the right Rindler wedge (R-region, $|t| < z$): $t= e^\xi \sinh \tau$, $z=e^{\xi}\cosh\tau$;
\item the left Rindler wedge (L-region $z < -|t|$): $t= e^{\tilde{\xi}}\sinh \tilde{\tau}$, $z = -e^{\tilde{\xi}}\cosh\tilde{\tau}$;
\item the future degenerate Kasner spacetime (F-region, $t > |z|$): $t= e^{\eta} \cosh \zeta$, $z = e^{\eta}\sinh \zeta$.
\end{itemize}
Thus, we are using the units such that the acceleration of the world line $\xi=0$ is $1$. (It would be straightforward to include the past degenerate
Kasner spacetime for the analysis in this section~\cite{HIUY}.)

Let us recall that the Klein-Gordon inner product between two complex solutions $f_A$ and $f_B$
of the scalar field equation, $(\nabla_\mu\nabla^\mu - m^2)f = 0$,
in the spacetime with the metric
\begin{eqnarray}
  ds^2=- N^2dt^2+ G_{ab}dx^a dx^b,
\end{eqnarray}
is defined by the following integral over the hypersurface with constant $t$:
\begin{eqnarray}
  && (f_A,f_B)_{\rm KG}=i\int d^3x \frac{\sqrt{G}}{N}
  (\overline{f_A}\partial_0 f_B-f_B\partial_0 \overline{f_A}).
  \label{KGnormap}
\end{eqnarray}
(See, e.g.\ Ref.~\cite{Higuchi}.) This inner product can readily be shown to be
time independent.
It is well known that the following mode functions defined in the F-region
form a complete set of positive-frequency solutions~\cite{BiD,diSessa,Sommerfield,FullingParkerHu}:
\begin{eqnarray}
u_p^F(x_F) & = & - \frac{i}{2\sqrt{2}}e^{\frac{\pi|p|}{2}} H_{i|p|}^{(2)}(me^{\eta})e^{-ip\zeta},
\label{future-sol-Mink}
\end{eqnarray}
where $p$ takes all real values.  Here $x_F = (\eta,\zeta)$, and the coordinates $\eta$ and $\zeta$ are the time and 
space coordinates, respectively.

The functions $u_p^F(x_F)$ can be shown to be superpositions of the positive-frequency solutions for the Minkowski vacuum as follows (see e.g., \cite{YST95}).
We use the integral representation for the Hankel function
\cite{Magnus}, 
\begin{equation}
e^{-i\alpha\cosh {\cal K}+i\beta\sinh {\cal K}}=
{1\over2i}\int_{-\infty}^{\infty}dp\, e^{-i{\cal K} p}
\Biggl({\alpha+\beta\over \alpha-\beta}\Biggr)^{ip/2}
e^{\pi p/2}H^{(2)}_{ip}\bigl((\alpha^2-\beta^2)^{1/2}\bigr),
\label{magnus-formula}
\end{equation}
with $\alpha=mt$, $\beta=mz$ and
${\cal K}={\rm arcsinh}(k/m)$, to find
\begin{equation}
e^{-i{\omega_k} t +ikz}={1\over 2i}\int_{-\infty}^{\infty}dp\,
e^{-i{\cal K} p} \Biggl({t+z\over t-z}\Biggr)^{ip/2}e^{\pi p/2}
H^{(2)}_{ip}\bigl(m(t^2-z^2)^{1/2}\bigr),
\label{eq:trfmb}
\end{equation}
where $\omega_k=m\cosh{\cal K}=\sqrt{k^2+m^2}$.
The variable $t$ has an infinitesimal negative imaginary part
so that the condition
${\rm Im}(\alpha\pm \beta)<0$ for the validity of Eq.~(\ref{magnus-formula}) is satisfied.
If we use the coordinates $(\eta,\zeta)$ introduced in the F-region, we have
\begin{eqnarray}
e^{-i{\omega_k} t +ikz} & = & {1\over 2i}\int_{-\infty}^{\infty}dp\,
e^{-i{\cal K} p} e^{ip\zeta}e^{\pi p/2} H^{(2)}_{ip}(me^{\eta}) \nonumber \\
& = &  \sqrt{2}\int_{-\infty}^{\infty}dp\,
e^{i{\cal K} p}u_p^F(x_F),  \label{linear-1}
\label{eq:trfmbb}
\end{eqnarray}
where we have used 
$e^{-\pi p/2}H^{(2)}_{-ip}(me^\eta) = e^{\pi p/2}H^{(2)}_{ip}(me^{\eta}) = e^{\pi |p|/2}H^{(2)}_{i|p|}(me^{\eta})$
and let $p \to - p$.
This equation can readily be inverted as
\begin{equation}
u_p^F(x_F) = \sqrt{2}\int_{-\infty}^\infty \frac{dk}{4\pi\omega_k}
\,e^{-i\mathcal{K}p}e^{-i\omega_k t + ikz},  \label{linear-2}
\end{equation}
where we have used $d\mathcal{K} = dk/\omega_k$.
Thus, the mode functions $u_p^F(x_F)$ are superpositions of the 
positive-frequency mode functions $e^{-i\omega_kt + ikz}$ for Minkowski vacuum and vice versa.
It is also well known that the functions $u^F_p(x_F)$ are normalized so that
\begin{eqnarray}
( u_{p}^F, u_{p'}^F)_{\textrm{KG}} = \delta^D(p-p').
\end{eqnarray}
As a result, if the quantum scalar field $\phi(x_F)$ is expanded as.
\begin{eqnarray}
  \phi & = &\int_{-\infty}^\infty\,
  dp\left[ u_p^F(x_F)\, \hat{b}_p + \overline{u_p^F(x_F)}\,\hat{b}_p^\dagger\right],
\label{original-expansion}
\end{eqnarray}
then one finds $[ \hat{b}_p, \hat{b}_{p'}^\dagger] = \delta^D(p - p')$ and $[\hat{b}_p,\hat{b}_{p'}] = 0$.  

The Minkowski vacuum state
$|0\rangle_\textrm{M}$ is defined by $\hat{b}_p|0\rangle_\textrm{M} = 0$ for all $p$. 
This fact can be
shown as follows.  If we expand the quantum scalar field $\phi$ in the standard way as
\begin{equation}
\phi = \int_{-\infty}^\infty \frac{dk}{\sqrt{4\pi\omega_k}}
\left( \hat{c}_k\,e^{-i\omega_k t + ikz} + \hat{c}_k^\dagger e^{i\omega_k t - ikz}\right),
\label{standard-expansion}
\end{equation}
then the Minkowski vacuum state is defined by the requirement that
$\hat{c}_k|0\rangle_\textrm{M} = 0$ for all $k$, as is well known.  By substituting Eq.~(\ref{linear-2}) into
Eq.~(\ref{original-expansion}) and comparing the result with Eq.~(\ref{standard-expansion}) we find
\begin{equation}
  \hat{c}_k = \frac{1}{\sqrt{2\pi\omega_k}}
  \int_{-\infty}^\infty dp\,e^{-i\mathcal{K}p}\,\hat{b}_p.
\end{equation}
On the other hand, if we substitute Eq.~(\ref{linear-1}) into Eq.~(\ref{standard-expansion}) and compare the result
with Eq.~(\ref{original-expansion}), we obtain
\begin{equation}
\hat{b}_p  = \int_{-\infty}^\infty \frac{dk}{\sqrt{2\pi \omega_k}}\,e^{i\mathcal{K}p}\,\hat{c}_k.
\end{equation}
Thus, the two conditions, (i) $\hat{c}_k|0\rangle_{\textrm{M}} = 0$ for all $k$ and (ii) 
$\hat{b}_p|0\rangle_{\textrm{M}} = 0$
for all $p$, are equivalent.  Hence, the Minkowski vacuum state can be defined by condition (ii) as well as by condition (i),
which is the standard one.

The analytic continuation of the positive-frequency modes $u_p(x_F)$ to the R- and L-regions is performed in such a way that
the Minkowski time coordinate $t$ has an infinitesimal negative imaginary part. (This is because the factor $e^{-i\sqrt{k^2 + m^2} t}$ for 
positive-frequency modes should be regularized
by letting $t \to t-i\epsilon$, $\epsilon>0$, so that any $k$-integration involving them converges for large $|k|$.)  As a result
$(t^2 - z^2)^\alpha = e^{2\alpha\eta}$ in the F-region
must be analytically continued to $e^{-i\pi\alpha}(z^2 -t^2)^\alpha = e^{2\alpha(\xi - \frac{i\pi}{2})}$ in the R-region, and similarly for the L-region.
This observation leads to the rules for analytic continuation given by Table~I.  We use this table and the formula 
\begin{eqnarray}
K_\nu (z) & = & - \frac{\pi i}{2}e^{-\frac{\pi}{2}\nu i}H_\nu^{(2)}(ze^{-\frac{\pi i}{2}}),\ \ - \frac{\pi}{2} < \textrm{arg}\,z \leq \pi,
\end{eqnarray}
with $z=m\xi$ or $m\tilde{\xi}$ and with $\nu = \pm ip$, to continue the solution $u^F(x_F)$ given by Eq,~(\ref{future-sol-Mink}) to the R- and L-regions.
Thus, we find the following results \cite{Fulling}:
\begin{eqnarray}
u^F_p(x_F) & \leftarrow & \begin{cases} u^{(+)}_p(x): = \dfrac{1}{\sqrt{1- e^{-2\pi|p|}}}\left(v^R_p(x) + e^{-\pi|p|}\overline{v^L_{p}(x)}\right)
& \textrm{for}\ \ p > 0, \\
u^{(-)}_{p}(x):= \dfrac{1}{\sqrt{1-e^{-2\pi |p|}}}\left( v^L_{|p|}(x) + e^{-\pi|p|}\overline{v^R_{|p|}(x)}\right)
& \textrm{for}\ \ p < 0, \end{cases}
\label{UF-continueAH}
\end{eqnarray}
where $x=x_R=(\tau,\xi)$ if $x$ is in the R-region and  $x=x_L = (\tilde{\tau},\tilde{\xi})$ if $x$ is in the L-region and where, for $p>0$,
\begin{eqnarray}
 v^R_p (x) & = & \begin{cases} \dfrac{\sqrt{\sinh(\pi|p|)}}{\pi}K_{ip}(m\xi )e^{-ip\tau} & \textrm{if}\ \ z> |t|\ \ (\textrm{R-region}),\\
0 & \textrm{if}\ \ z < -|t|\ \ (\textrm{L-region}),\end{cases} \\
v^L_p(x) & = & \begin{cases} 0 & \textrm{if}\ \ z > |t|\ \ (\textrm{R-region}),\\
\dfrac{\sqrt{\sinh(\pi|p|)}}{\pi}K_{ip}(m\tilde{\xi})e^{-ip\tilde{\tau}} & \textrm{if}\ \ z < -|t|\ \ (\textrm{L-region}). \end{cases}
\end{eqnarray}
(We have made the arrow left-pointing to indicate the direction of time evolution.)
 Note that, in each of these regions, the coordinates $\tau$ ($\tilde{\tau}$) and
$\xi$ ($\tilde{\xi}$) are the time and space coordinates, respectively.

It is interesting to find the mode functions in the F-region
that are obtained by evolving the right and left Rindler modes, $v_p^R(x)$ and $v_p^L(x)$.  It is important to
note that they are not obtained by analytic continuation because $v_p^R(x)$ and $v_p^L(x)$ are not analytic functions, being $0$ on
open regions.  Instead we use Eq.~(\ref{UF-continueAH}) and its complex conjugate in reverse with the observation that, with $p>0$,
\begin{eqnarray}
v^R_p(x) & = & \frac{1}{\sqrt{1-e^{-2\pi p}}}\left[u_p^{(+)}(x) - e^{-\pi p}\overline{u_p^{(-)}(x)}\right]
\to v^F_p (x_F) := \frac{1}{\sqrt{1 -e^{-2\pi p}}}\left[ u_p^F(x_F) - e^{-\pi p} \overline{u_{-p}^F(x_F)}\right] ,  \label{defvF1} \\
v^L_p(x) &  = & \frac{1}{\sqrt{1-e^{-2\pi p}}}\left[ u_p^{(-)}(x) - e^{-\pi p}\overline{u_p^{(+)}(x)}\right]
\to v^F_{-p}(x_F) := \frac{1}{\sqrt{1 - e^{-2\pi p}}}\left[ u_{-p}^F(x_F) - e^{-\pi p}\overline{u_p^F(x_F)}\right] . \label{defvF2}
\end{eqnarray}
Thus, we obtain~\cite{Sommerfield}
\begin{eqnarray}
v_p^F(x_F) &  = & - \frac{i}{2\sqrt{\sinh (\pi |p|)}}J_{-i|p|}(me^{\eta}) e^{-ip\zeta},\label{vp}
\end{eqnarray}
where we have used
\begin{eqnarray}
J_{-i|p|}(y) = \frac{1}{2}\left[ e^{\pi|p|}H_{i|p|}^{(2)}(y) + \overline{H_{i|p|}^{(2)}(y)}\right] \ \ \textrm{for}\ y\ \textrm{real}.
\end{eqnarray}
It is clear from Eqs.~(\ref{defvF1}) and (\ref{defvF2}) that the modes $v_p^F(x_F)$, which we call the Kasner modes, satisfy
$( v_p^F, v_{p'}^F)_{\textrm{KG}} = \delta^D(p-p')$ for all real $p$ and $p'$. (Note that
$(\overline{u_p^F},\overline{u_{p'}^F})_{\textrm{KG}} = - \delta^D(p-p')$ and that
$( \overline{u_p^F}, u_{p'}^F)_{\textrm{KG}} = 0$.)  This fact can directly be verified from Eq.~(\ref{vp}).
First we note, using the metric,
\begin{eqnarray}
ds^2 = e^{2\eta}\left( - d\eta^2 + d\zeta^2\right),
\end{eqnarray}
that the Klein-Gordon inner product between these modes is given by
\begin{eqnarray}
( v_p^F, v_{p'}^F)_{\textrm{KG}} = i
\int_{-\infty}^\infty d\zeta \left[ \overline{v_p^F(\eta,\zeta)}\frac{\partial\ }{\partial\eta}v_{p'}^F(\eta,\zeta)
- \frac{\partial\ }{\partial \eta} \overline{v_p^F(\eta,\zeta)}\cdot v_{p'}^F(\eta,\zeta)\right].
\end{eqnarray}
 The integral can readily be evaluated by noting
that, near the horizon, i.e.\ for large and negative $\eta$, we have from Eq.~(\ref{vp})
\begin{eqnarray}
v_p^F(x_F) & \approx & - \frac{i}{2\sqrt{\sinh(\pi |p|)}}\cdot\frac{1}{\Gamma(1- i|p|)}\left( \frac{m e^\eta}{2}\right)^{-i|p|}e^{-ip\zeta} \nonumber \\
& =  & \frac{1}{\sqrt{4\pi |p|}}e^{-i|p|\eta - ip\zeta + i\delta_v},
\end{eqnarray}
where
\begin{eqnarray}
e^{i\delta_v} & = & -i \left[ \frac{\Gamma(1+ip)}{\Gamma(1-ip)}\right]^{\frac{1}{2}}\left( \frac{m}{2}\right)^{-ip}.
\end{eqnarray}
We have used
\begin{eqnarray}
|\Gamma(1-ip)|^2 = \frac{\pi|p|}{\sinh(\pi|p|)}.
\end{eqnarray}
Notice that the modes $v_p^F(x_F)$ are purely left-moving if $p>0$ whereas they are purely right-moving if $p< 0$.

Now we can use the relations (\ref{UF-continueAH}) or (\ref{defvF1}) and (\ref{defvF2})  to examine the entanglement structure of the
Minkowski vacuum state $|0\rangle_\textrm{M}$.   We expand the scalar field $\phi(x_F)$ in the F-region in terms of $v_p^F(x_F)$ as
\begin{eqnarray}
\phi(x_F) & = & \int_{-\infty}^\infty dp\left[ v_p^F(x_F) \hat{a}_p + \overline{v_p^F(x_F)}\hat{a}_p^\dagger\right]. \label{future-phi-v}
\label{phi-in-F}
\end{eqnarray}
By Eqs.~(\ref{defvF1}) and (\ref{defvF2}), in the union of the R- and L-regions Eq.~(\ref{phi-in-F})  becomes
\begin{eqnarray}
\phi(x) & = & \int_{0}^\infty dp\left[ v_p^R(x)\hat{a}_p + v_p^L(x)\hat{a}_{-p}
+ \overline{v_p^R(x)}\hat{a}_p^\dagger + \overline{v_p^L(x)}\hat{a}_{-p}^\dagger\right].
\end{eqnarray}
By comparing Eq.~(\ref{original-expansion})
with the expression obtained by substituting Eqs.~(\ref{defvF1}) and (\ref{defvF2}) into Eq.~(\ref{future-phi-v}), one finds
\begin{eqnarray}
\hat{b}_p & = & \frac{1}{\sqrt{1- e^{-2\pi |p|}}}\left( \hat{a}_p - e^{-\pi|p|}\hat{a}_{-p}^\dagger\right),
\end{eqnarray}
for all $p$.
Thus, the condition on the Minkowski vacuum state, $\hat{b}_p|0\rangle_{\textrm{M}} = 0$ for all $p$, becomes
\begin{eqnarray}
(\hat{a}_p - e^{-\pi|p|}\hat{a}_{-p}^\dagger)|0\rangle_\textrm{M} & = & 0.
\label{condition-on-M}
\end{eqnarray}
 Then, defining the Rindler vacuum state $|0\rangle_{\textrm{Rin}}$ by the conditions
$\hat{a}_p|0\rangle_{\textrm{Rin}} = 0$ for all $p$, one finds that condition (\ref{condition-on-M}) leads to
\begin{eqnarray}
|0\rangle_{\textrm{M}} = \mathcal{N} \exp\left( \int_0^\infty dp\,e^{-\pi p}\hat{a}_p^\dagger \hat{a}_{-p}^\dagger\right)|0\rangle_{\textrm{Rin}},
\label{Mink-Rind}
\end{eqnarray}
where $\mathcal{N}$ is an (infinitesimal) normalization factor.  Heuristically, if we approximate this formula by discretizing the momentum so that
$[\hat{A}_p,\hat{A}_{p'}^\dagger] = \delta_{pp'}$, where $A_p \propto a_p$, then it is approximated as
\begin{eqnarray}
|0\rangle_{\textrm{M}} & = & \mathcal{N} \prod_{p>0} \sum_{n=0}^\infty \frac{e^{-\pi p n}}{n!} (A_p^\dagger)^n (A_{-p}^\dagger)^n
|0\rangle_{\textrm{Rin}} \nonumber \\
& = & \mathcal{N} \prod_{p>0} \sum_{n=0}^\infty e^{-\pi p n} |p, n\rangle_R \otimes |p, n\rangle_L,
\label{M-entangle}
\end{eqnarray}
where $|p,n\rangle_R$ ($|p,n\rangle_L$) are the $n$-particle states of the right (left) Rindler mode $v_p^{R}(x)$ ($v_p^L(x)$).
Thus, as is well known, the Minkowski vacuum state is an entangled state in terms of the left and right Rindler states.  One can also consider it as a
state with entanglement between the states corresponding to the modes $v_p^F(x_F)$ and $v_{-p}^F(x_F)$ in the F-region.

\begin{figure}[t]
  \begin{center}
        \includegraphics[width=8cm]{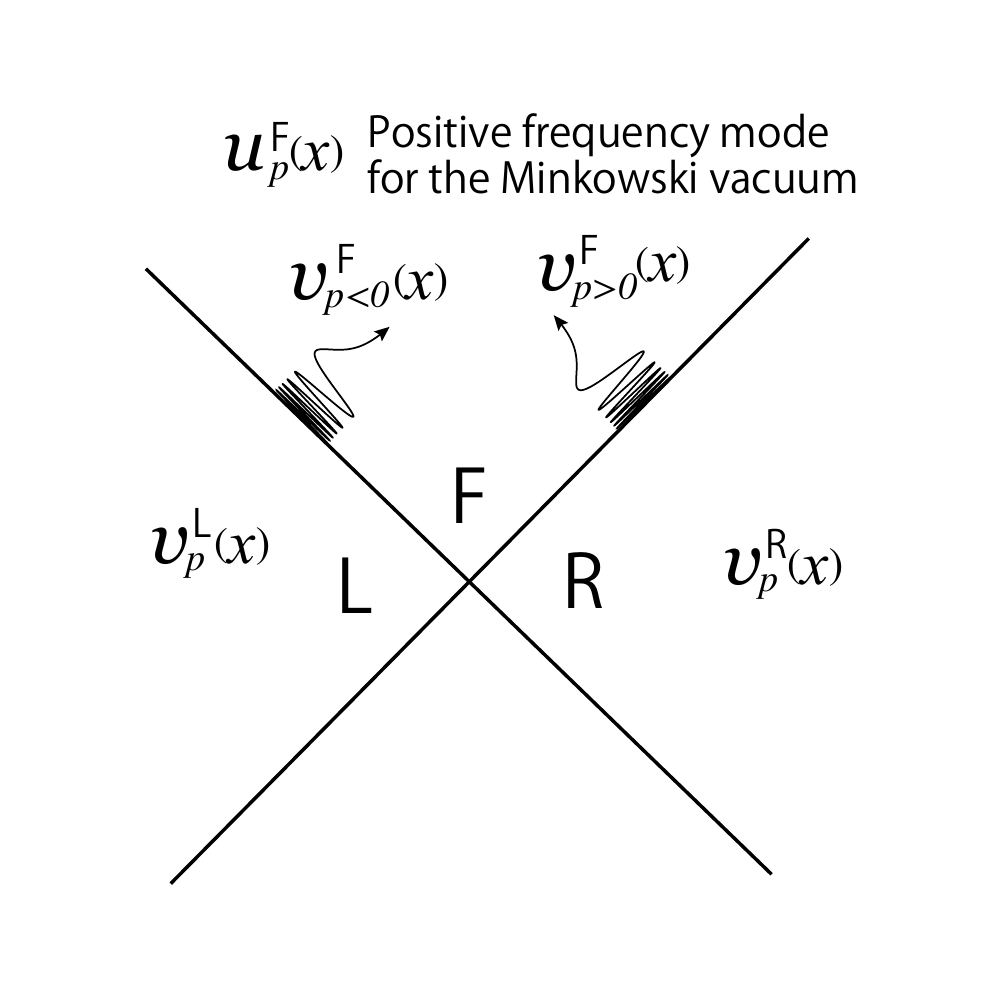}%
  \end{center}
\caption{
  \label{spfigure} Relations among modes in the R-, F- and  L-regions.
  The functions $v_p^R$ and $v_p^L$ are the positive-frequency mode functions
  in the R-region and the L-region, respectively.
  The function $u_p^F$ is the positive-frequency mode function of the global spacetime,
    which is obtained by the analytic continuation from the R-region to the F-region.
    The functions $v_{p>0}^F$ and $v_{p<0}^F$ denote mode functions in the F-region,
    which behave as the left-moving wave mode and the right-moving wave mode,
    respectively, near the
    horizon. The modes $v_{p>0}^F$ and $v_{p<0}^F$ continued into
    the R-region and the L-region yield $v_p^R$ and $v_{-p}^L$, respectively.
    The analytic continuation of $v_p^R$ into the L-region yields $e^{-\pi p}\overline{v_p^L}$.}
\end{figure}

\section{Various charts of de~Sitter spacetime} \label{sec:three}
In this section we list coordinates for various charts of de~Sitter spacetime used in this paper.
This spacetime is the maximally symmetric  solution of the
Einstein equations with a positive cosmological constant $\Lambda=3H^2$.
It can be
described as a $4$-dimensional hypersurface in the 5-dimensional Minkowski
spacetime with cartesian coordinates $z^\mu$, $\mu=0,1,2,3,4$, with the line element
\begin{eqnarray}
  &&ds^2=-(dz^0)^2+(dz^1)^2+(dz^2)^2+(dz^3)^2+(dz^4)^2. \label{dS-line}
\end{eqnarray}
The condition that gives de~Sitter spacetime is
\begin{eqnarray}
  &&-(z^0)^2+(z^1)^2+(z^2)^2+(z^3)^2+(z^4)^2={1\over H^2}.
\end{eqnarray}
We adopt the units such that $H=1$ from now on.
\begin{itemize}
\item The right conformally-flat chart describes the region with $-z^1 < z^0$.  The coordinates are $\eta_R$, $r_R$, $\theta$ and $\varphi$, where
\begin{eqnarray}
&& z^0 = \frac{1}{2}\left( - \frac{1}{\eta_R} + \eta_R - \frac{r_R^2}{\eta_R}\right), \ \
z^1 = \frac{1}{2}\left( - \frac{1}{\eta_R} - \eta_R +\frac{r_R^2}{\eta_R} \right),\ \
z^j = - \frac{r_R}{\eta_R}\hat{n}^j,\ \ j=2,3,4,
\end{eqnarray}
with $(\hat{n}^2,\hat{n}^3,\hat{n}^4) = (\cos\theta,\sin\theta\cos\varphi,\sin\theta\sin\varphi)$, $-\infty< \eta_R < 0$ and $0 \leq r_R$.
\item The left conformally-flat chart describes the region with $z^1 < z^0$.  The coordinates are $\eta_L$, $r_L$, $\theta$ and $\varphi$, where
\begin{eqnarray}
&& z^0 = \frac{1}{2}\left( - \frac{1}{\eta_L} + \eta_L - \frac{r_L^2}{\eta_L}\right), \ \
z^1 = \frac{1}{2}\left( \frac{1}{\eta_L} + \eta_L - \frac{r_L^2}{\eta_L} \right),\ \
z^j = - \frac{r_L}{\eta_L}\hat{n}^j,\ \ j=2,3,4,
\end{eqnarray}
with $- \infty < \eta_L < 0$ and $0 \leq r_L$.
\item The right static chart describes the region where $|z^0| < z^1$ (the R-region).  The coordinates are $T_A$, $R_A$,  $\theta$ and $\varphi$, where
\begin{eqnarray}
&& z^0 = \sqrt{1-R_A^2}\sinh T_A, \ \
z^1 = \sqrt{1- R_A^2}\cosh T_A,\ \
z^j = R_A \hat{n}^j,\ \ j=2,3,4,
\end{eqnarray}
with $-\infty < T_A < \infty$ and  $0 \leq R_A <1$.  This chart is part of the right conformally-flat chart and their coordinates are related by
\begin{eqnarray}
&& R_A = - \frac{r_R}{\eta_R},\ \ e^{-T_A} = \sqrt{\eta_R^2 - r_R^2}.
\end{eqnarray}
\item The left static chart describes the region where $z^1 < - |z^0|$ (the L-region).  The coordinates are $T_L$, $R_L$, $\theta$ and $\varphi$, where
\begin{eqnarray}
&& z^0 = \sqrt{1-R_L^2}\sinh T_L, \ \
z^1 = - \sqrt{1- R_L^2}\cosh T_L,\ \
z^j = R_L \hat{n}^j,\ \ j=2,3,4,
\end{eqnarray}
with  $- \infty < T_L < \infty$ and $0 \leq R_L <1$.  This chart is part of the left conformally-flat chart and their coordinates are related by
\begin{eqnarray}
&& R_L= - \frac{r_L}{\eta_L},\ \ e^{-T_L} = \sqrt{\eta_L^2 - r_L^2}.
\end{eqnarray}
\item The future region is the region where $z^0 > |z^1|$ (the F-region).  The coordinates are $R_B$, $T_B$, $\theta$ and $\varphi$, where
\begin{eqnarray}
&& z^0 = \sqrt{R_B^2- 1}\cosh T_B, \ \
z^1 = \sqrt{R_B^2- 1}\sinh T_B,\ \
z^j = R_B \hat{n}^j,\ \ j=2,3,4,
\end{eqnarray}
with $1 < R_B < \infty$ and $-\infty < T_B < \infty$.  This region is the intersection of the left and right conformally-flat charts.  The coordinates for these charts
are related as follows:
\begin{eqnarray}
&& R_B = - \frac{r_R}{\eta_R} = - \frac{r_L}{\eta_L},\ \
e^{-T_B} = \sqrt{r_R^2 - \eta_R^2} = \frac{1}{\sqrt{r_L^2 - \eta_L^2}}. \label{F-R-L}
\end{eqnarray}
The line element (\ref{dS-line}) is given in the left or right conformally-flat chart as
\begin{eqnarray}
ds^2  & = & \frac{1}{\eta^2}(-d\eta^2 + dr^2 + r^2 d\Omega_{(2)}^2),
\end{eqnarray}
where $(r,\eta) = (r_R,\eta_R)$ or $(r_L,\eta_L)$.  The line element of the unit $2$-sphere is denoted by $d\Omega_{(2)}^2$ here.  In the
left or right static chart, i.e.\ in the R- or L-region, it is given by
\begin{eqnarray}
ds^2 & = & -(1-R^2)dT^2 + \frac{dR^2}{1-R^2} + R^2d\Omega_{(2)}^2,
\end{eqnarray}
where $(T,R) = (T_A,R_A)$ or $(T_L,R_L)$.  Finally, in the F-region, we find
\begin{eqnarray}
ds^2 & = &  - \frac{dR_B^2}{R_B^2 - 1} + (R_B^2 -1)dT_B^2 + R_B^2 d\Omega_{(2)}^2.  \label{F-metric}
\end{eqnarray}
\end{itemize}

\def\r{\rho}

In this paper
we study the relationship between solutions of the scalar field equation, $(\nabla_\mu \nabla^\mu - m^2)\phi = 0$, in these charts.  In any of these charts the
line element takes the form,
\begin{eqnarray}
&& ds^2 = - [N(\r,t)]^2 dt^2 +[M(\r,t)]^2 d\r^2 + [K(\r,t)]^2 d\Omega_{(2)}^2.
\end{eqnarray}
In all cases the complete set of solutions to the scalar field equation can be given as
\begin{eqnarray}
&& \psi_{\kappa \ell m}(t,\r,\theta,\varphi)  =  \varphi_{\kappa\ell}(t,\r) Y_{\ell m}(\theta,\varphi),
\end{eqnarray}
and $\overline{\psi_{\kappa\ell m}(t,\r,\theta,\varphi)}$, where $\kappa$ is a continuous label.  The Klein-Gordon inner product is
\begin{eqnarray}
(\psi_{\kappa\ell m}, \psi_{\kappa'\ell' m'})_{\textrm{KG}}
& = &  - i \int d\r d\theta d\varphi\,\sqrt{-g}\left[ \overline{\psi_{\kappa \ell m}}\partial^t \psi_{\kappa'\ell'm'}
- \partial^t \overline{\psi_{\kappa \ell m}}\cdot \psi_{\kappa'\ell'm'}\right].
\end{eqnarray}
By the orthonormality of the spherical harmonics $Y_{\ell m}(\theta,\varphi)$ we find
\begin{eqnarray}
(\psi_{\kappa\ell m},\psi_{\kappa'\ell'm'})_{\textrm{KG}} = (\varphi_{\kappa\ell},\varphi_{\kappa'\ell})_{\textrm{RKG}}\delta_{\ell\ell'}\delta_{mm'}.
\end{eqnarray}
where $(\bullet,\bullet)_{\textrm{RKG}}$ is the radial Klein-Gordon inner product defined by
\begin{eqnarray}
(\varphi_{\kappa\ell},\varphi_{\kappa'\ell})_{\textrm{RKG}}
& = & i \int d\r \frac{M(\r,t)[K(\r,t)]^2}{N(\r,t)}
\left[ \overline{\varphi_{\kappa\ell}}\frac{\partial\varphi_{\kappa'\ell}}{\partial t}
- \frac{\partial\overline{\varphi_{\kappa\ell}}}{\partial t}\varphi_{\kappa'\ell}\right].
\end{eqnarray}
Thus, if we  normalize the functions $\varphi_{\kappa\ell}(\r,t)$, which we call the temporal-radial part of $\psi_{\kappa\ell m}(t,\r,\theta,\varphi)$, by
$(\varphi_{\kappa\ell},\varphi_{\kappa'\ell})_{\textrm{RKG}} = \delta^D(\kappa - \kappa')$ [and
$(\varphi_{\kappa\ell},\overline{\varphi_{\kappa'\ell}})_{\textrm{RKG}} = 0$], and if the quantum field $\phi(t,\r,\theta,\varphi)$ is expanded as
\begin{eqnarray}
\phi(t,\r,\theta,\varphi) = \int d\kappa \sum_{\ell m} \left[ \psi_{\kappa\ell m}(t,\r,\theta,\varphi) \hat{d}_{\kappa\ell m}
+ \overline{\psi_{\kappa\ell m}(t,\r,\theta,\varphi)}\hat{d}_{\kappa\ell m}^\dagger \right],
\end{eqnarray}
then $[\hat{d}_{\kappa\ell m},\hat{d}_{\kappa'\ell'm'}^\dagger] = \delta^D(\kappa-\kappa')\delta_{\ell\ell'}\delta_{mm'}$
with all other commutators vanishing.

\section{Positive-frequency solutions for the Bunch-Davies vacuum in the future region} \label{sec:four}
In this section we present the positive-frequency modes that are natural to the F-region for the scalar field of mass $m$ with no coupling to the scalar curvature
for the Bunch-Davies vacuum.
It is well known that
a complete set of positive-frequency modes for the Bunch-Davies vacuum is given in the right conformally-flat chart  by
\begin{eqnarray}
&& \psi_{k\ell m}(\eta_R,r_R,\theta,\varphi) = \varphi_{k\ell}(r_R,\eta_R)Y_{\ell m}(\theta,\varphi),
\end{eqnarray}
where
\begin{eqnarray}
&& \varphi_{k\ell}(r_R,\eta_R) = \frac{e^{-\frac{i\pi}{2}\left(\ell+\frac{1}{2}\right)}}{\sqrt{2k}}
(-k\eta_R)^{\frac{3}{2}}e^{\frac{i\nu\pi}{2}}H_\nu^{(1)}(-k\eta_R)j_\ell(kr_R),
\ \ k > 0,  \label{def-varphi}
\end{eqnarray}
with
\begin{eqnarray}
\nu = \sqrt{\frac{9}{4} - m^2}.
\label{value-of-nu}
\end{eqnarray}
Thus, if we expand the quantum scalar field $\phi$ as
\begin{equation}
\phi = \int_0^\infty dk\,\sum_{\ell,m} \left[ \psi_{k\ell m}(\eta_R,r_R,\theta,\varphi)\hat{c}_{k\ell m}
+ \overline{\psi_{k\ell m}(\eta_R,r_R,\theta,\varphi)}\hat{c}_{k\ell m}^\dagger\right],
\label{standard-BD-expansion}
\end{equation}
then the Bunch-Davies vacuum state $|0\rangle_{\textrm{BD}}$ is defined by requiring that
$\hat{c}_{k\ell m}|0\rangle_{\textrm{BD}}=0$ for all $k$, $\ell$ and $m$.
It is also well known that they satisfy $(\varphi_{k\ell},\varphi_{k'\ell})_{\textrm{RKG}} = \delta^D(k-k')$ 
(see, e.g.\ Ref.~\cite{HiguchiA}).
By using the formula~\cite{Magnus}\footnote{
    Originally, the mathematical formula is given with the use of $K_\nu(ax)$ instead of
    $H_\nu^{(1)}(az)$, but they are related by $K_\nu(ax)=\pi i e^{\pi \nu i/2} H_\nu^{(1)}(iax)/2$. }
\begin{eqnarray}
  \int_0^\infty dz z^\lambda H_\nu^{(1)}(az) J_\mu(bz)&=&a^{-\lambda-1}e^{(\lambda-\nu+\mu)i\pi/2}
      {2^\lambda (b/a)^\mu \over\pi \Gamma(\mu+1)}\Gamma\biggl({\lambda+\nu+\mu+1\over2}\biggr)
      \Gamma\biggl({\lambda-\nu+\mu+1\over2}\biggr)
      \nonumber\\
          &&{}_2F_1\biggl(
      {\lambda+\nu+\mu+1\over2}, {\lambda-\nu+\mu+1\over2},\mu+1;(b/a)^2 \biggr),
\label{HJml}
\end{eqnarray}
which is valid for ${\rm Re}(-ia\pm ib)>0, ~{\rm Re}(\mu+\lambda+1\pm\nu)>0$, 
with $\lambda = - ip$, $\mu= \ell +\frac{1}{2}$, $a=-\eta$, $b=r$ and with $\nu$ given by 
Eq.~(\ref{value-of-nu}) and with the assumption that
$r_R < -\eta_R$ for now, we find
\begin{eqnarray}
\frac{1}{\sqrt{2\pi}}\int_0^\infty dk\, k^{-ip-\frac{1}{2}}\varphi_{k\ell}(\eta_R,r_R)
& = & \frac{2^{-ip}}{2\sqrt{2}\pi\Gamma(\ell+\frac{3}{2})}
   \Gamma\biggl({\frac{3}{2}+\ell-ip+\nu\over2}\biggr)
      \Gamma\biggl({\frac{3}{2}+\ell-ip-\nu\over2}\biggr) \nonumber \\
&& \times
e^{\frac{\pi p}{2}}\left( (-\eta_R)^2 - r_R^2\right)^{\frac{ip}{2}}U^{(S)}_{|p|\ell}(-r_R/\eta_R),  \label{pos-freq-F}
\label{desitterpositive}
\end{eqnarray}
where
\begin{eqnarray}
U^{(S)}_{|p|\ell}(x) := x^{\ell} (1 - x^2)^{\frac{i|p|}{2}}
 {}_2F_1\biggl(
          {\frac{3}{2}+\ell+i|p|+\nu\over2}, {\frac{3}{2} + \ell + i|p|-\nu\over2},\ell+{3\over 2}; x^2 \biggr).
\label{US-def-ah}
\end{eqnarray}
We have used the formula
\begin{eqnarray}
&& {}_2F_1(a,b,c;z) = (1-z)^{c-a-b}{}_2F_1(c-a,c-b,c;z)
\end{eqnarray}
for $p > 0$.  This formula can also be used to show that the function $U_{|p|\ell}(x)$ is real if $0 \leq x < 1$. 
Note that the variable $\eta_R$ must have an infinitesimal negative imaginary part for the validity of 
Eq.~(\ref{desitterpositive}) because of the condition ${\rm Re}(-ia\pm ib)>0$ for Eq.~(\ref{HJml}).

Since the function $\varphi_{k\ell}(\eta_R,r_R)$ are (the temporal-radial part of) positive-frequency modes
for the Bunch-Davies vacuum, so is the right-hand side of Eq.~(\ref{pos-freq-F}) (in the R-region).
In order to find the expression for this function in the F-region, we need to examine how it should be analytically continued from
$r_R < -\eta_R$ (R-region)  to $r_R > -\eta_R$ (F-region).  To do so, 
we use the fact that the variable $\eta_R$ must have an infinitesimal negative imaginary part 
for the integral in Eq.~(\ref{pos-freq-F}), as we mentioned above,
so that the $k$-integral converges.
This implies in particular that
$[1 - (-r_R/\eta_R)^2]^\alpha$ with $r_R < -\eta_R$ should be continued to 
$e^{i\pi \alpha}[(-r_R/\eta_R)^2 - 1]^\alpha$.  By performing this
analytic continuation in Eq.~(\ref{pos-freq-F}) and using Eq.~(\ref{F-R-L}) we find
\begin{eqnarray}
 \frac{1}{\sqrt{2\pi}}\int_0^\infty dk\,k^{-ip-\frac{1}{2}} \varphi_{k\ell}(\eta_R,r_R) = 2^{-ip}e^{i\delta_p} u_{p\ell}^F(R_B,T_B),
\label{u-F}
\end{eqnarray}
where
\begin{eqnarray}
u_{p\ell}^F(R_B,T_B) & = & e^{-\frac{\pi|p|}{2}}N_{p\ell} U_{|p|\ell}(R_B)e^{-ipT_B}, \label{def-of-upl}
\end{eqnarray}
with
\begin{eqnarray}
N_{p\ell} & = &  \frac{1}{2\sqrt{2}\pi\Gamma(\ell+\frac{3}{2})}
\left|   \Gamma\biggl({\frac{3}{2}+\ell-ip+\nu\over2}\biggr)
\Gamma\biggl({\frac{3}{2}+\ell-ip-\nu\over2}\biggr)\right|,
\end{eqnarray}
and
\begin{eqnarray}
&& e^{i\delta_p}  =  \left[  \frac{\Gamma\biggl({\frac{3}{2}+\ell-ip+\nu\over2}\biggr)
\Gamma\biggl({\frac{3}{2}+\ell-ip-\nu\over2}\biggr)}
{ \Gamma\biggl({\frac{3}{2}+\ell+ip+\nu\over2}\biggr)
\Gamma\biggl({\frac{3}{2}+\ell+ip-\nu\over2}\biggr)}\right]^{1/2}.
\end{eqnarray}
The label $p$ can take any real value. Note that the coordinate $T_B$ is a spatial coordinate in the F-region [see Eq.~(\ref{F-metric})].
Here, we have defined
\begin{eqnarray}
U_{|p|\ell}(R_B) & = & R_B^\ell (R_B^2 -1)^{\frac{i|p|}{2}}
 {}_2F_1\biggl(
          {\frac{3}{2}+\ell+i|p|+\nu\over2}, {\frac{3}{2} + \ell + i|p|-\nu\over2},\ell+{3\over 2}; R_B^2 \biggr).
\label{U-def-ah}
\end{eqnarray}
Thus we have shown 
that Eq.~(\ref{u-F}) can be derived from the mathematical formula (\ref{HJml})
with the assumption that $r_R > -\eta_R$, i.e.\ in the F-region.
(The mode functions $u^F_{p\ell}(R_B,T_B)$ are proportional to those found by
Markkanen~\cite{Markkanen} for $m^2 = 2$, but our normalization factor
is different from his.\footnote{
Our normalization factor disagrees with that of Eq. (95) in Ref.~\cite{Markkanen} by
a factor of $1/\sqrt{1-e^{-2\pi k/H}}$.})
One can also invert Eq.~(\ref{u-F}) as
\begin{eqnarray}
\varphi_{k\ell}(\eta_R,r_R) = \frac{1}{\sqrt{2\pi k}}\int_{-\infty}^\infty dp\,k^{ip}2^{-ip}e^{i\delta_p}
u_{p\ell}^F(R_B,T_B),
\label{invert-BD}
\end{eqnarray}
by using
\begin{eqnarray}
&& \frac{1}{2\pi\sqrt{kk'}}\int_{-\infty}^\infty dp\left( \frac{k}{k'}\right)^{ip} = \delta^D(k'-k).
\end{eqnarray}
Thus, the functions $u_{p\ell}^F(x_F)Y_{\ell m}(\theta,\varphi)$ are superpositions of
the positive-frequency solutions 
$\psi_{k\ell m}(\eta_R,r_R,\theta,\varphi) = \varphi_{k\ell}(\eta_R,r_R)Y_{\ell m}(\theta,\varphi)$ for the
Bunch-Davies vacuum, and vice versa.

Let us expand the field in the F-region as
\begin{eqnarray}
  \phi(x_F,\theta,\varphi) & = & \sum_{\ell, m}\int_{-\infty}^\infty
  dp\left[ u^F_{p\ell}(x_F)Y_{\ell m}(\theta,\varphi)\,\hat{b}_{p\ell m}
+ \overline{u^F_{p\ell}(x_F)Y_{\ell m}(\theta,\varphi)}\,\hat{b}^\dagger_{p\ell m}\right].
\label{F-region-expansion}
\end{eqnarray}
Then, Eqs.~(\ref{u-F}) and (\ref{invert-BD}) imply, as in the Minkowski case, that the condition for the Bunch-Davies
vacuum state, $\hat{c}_{k\ell m}|0\rangle_{\textrm{BD}}=0$ for all $k$, $\ell$ and $m$, can be given as
$\hat{b}_{p\ell m}|0\rangle_{\textrm{BD}} = 0$ for all $p$, $\ell$ and $m$.

We note that, since the field equation depends on $\ell$ only through $\ell(\ell+1)$, which is invariant under $\ell \leftrightarrow - \ell-1$, another solution
can be obtained by replacing $U_{|p|\ell}(R_B,T_B)$ in the expression of $u^F_{p\ell}(R_B,T_B)$ by
\begin{eqnarray}
W_{|p|\ell}(R_B) & = & R_B^{-\ell-1} (R_B^2 -1)^{\frac{i|p|}{2}}
 {}_2F_1\biggl(
          {\frac{1}{2}-\ell+i|p|+\nu\over2}, {\frac{1}{2} - \ell + i|p|-\nu\over2},\frac{1}{2} - \ell; R_B^2 \biggr).
\end{eqnarray}

We find  $(u_{p\ell}^F, u_{p'\ell}^F)_{\textrm{KG}} = \delta^D(p-p')$ from  Eq.~(\ref{u-F}) by using
$(\varphi_{k\ell},\varphi_{k'\ell})_{\textrm{KG}} = \delta^D(k-k')$ and the formula
\begin{eqnarray}
&& \frac{1}{2\pi} \int_0^\infty \frac{dk}{k}\,k^{i(p'-p)} = \delta^D(p - p').
\end{eqnarray}
We can express (the temporal-radial part of) the mode functions, $u^F_{p\ell}(R_B,T_B)$, in terms of the positive-frequency
modes $\varphi_{p\ell}(\eta_L,r_L)$ in the left conformally-flat chart in exactly the same way.  Thus, we have
\begin{eqnarray}
 \frac{1}{\sqrt{2\pi}}\int_0^\infty dk\,k^{ip-\frac{1}{2}} \varphi_{k\ell}(\eta_L,r_L) = 2^{ip}e^{i\delta_{-p}} u_{p\ell}^F(R_B,T_B).
\label{u-F-2}
\end{eqnarray}

Finally, let us emphasize that the analytic continuation from $R_B < 1$ to $R_B>1$
of the hypergeometric function in the definition (\ref{U-def-ah}) of $U_{|p|\ell}(R_B)$ is
not unique and that it had to be specified as described before Eq.~(\ref{u-F}).
The function $U_{|p|\ell}(x)$  is not real for $x>1$ although it is for $0\leq x <1$.  These facts can be made clearer by expressing
it in terms of hypergeometric functions with argument $1-R_B^2$ with the use of the following formula:
\begin{eqnarray}
  {}_2F_1\left(a,b,c;z\right) &  = & \frac{\Gamma(c)\Gamma(c-a-b)}{\Gamma(c-a)\Gamma(
    c-b)}{}_2F_1\left(a,b,a+b-c+1;1-z\right)\nonumber \\
  && + (1-z)^{c-a-b}\frac{\Gamma(c)\Gamma(a+b-c)}{\Gamma(a)\Gamma(b)}{}_2F_1\left(c-a,c-b,c-a-b+1;1-z\right).
\end{eqnarray}
Thus, we find
\begin{eqnarray}
  U_{|p|\ell}(R_{B})&=&  \frac{\Gamma(\ell+\frac{3}{2})\Gamma(i|p|)}{\Gamma\left(\frac{\frac{3}{2}+\ell+i|p|+\nu}{2}\right)
\Gamma\left(\frac{\frac{3}{2}+\ell+i|p|-\nu}{2}\right)}e^{\pi|p|}V_{|p|\ell}(R_B)
 +  \frac{\Gamma(\ell+\frac{3}{2})\Gamma(-i|p|)}{\Gamma\left(\frac{\frac{3}{2}+\ell-i|p|+\nu}{2}\right)
\Gamma\left(\frac{\frac{3}{2}+\ell-i|p|-\nu}{2}\right)}\overline{V_{|p|\ell}(R_B)},  \label{UV}
\end{eqnarray}
where
\begin{eqnarray}
V_{|p|\ell}(R_B) & = & R_B^\ell (R_B^2 - 1)^{-\frac{i|p|}{2}}
 {}_2F_1\biggl(
          {\frac{3}{2}+\ell- i|p|+\nu\over2}, {\frac{3}{2} + \ell - i|p|-\nu\over2},1- i|p|; 1-  R_B^2 \biggr).
\end{eqnarray}
There is no ambiguity in the hypergeometric function here because it is analytic for all positive $R_B$.  The relation (\ref{UV}) can readily be inverted as
\begin{eqnarray}
V_{|p|\ell}(R_B) &  = &  \frac{\Gamma\left(\frac{\frac{3}{2}+\ell+i|p|+\nu}{2}\right)
\Gamma\left(\frac{\frac{3}{2}+\ell+i|p|-\nu}{2}\right)}{2\sinh(\pi|p|)\Gamma(\ell+\frac{3}{2})\Gamma(i|p|)}
\left[ U_{|p|\ell}(R_B) - e^{-\pi|p|}\overline{U_{|p|\ell}(R_B)}\right].  \label{UV-reverse}
\end{eqnarray}

\begin{center}
\begin{table}[t]
\begin{tabular}{llll}
  \hline
  \hline\\
~~~~~~
 ${\rm F} \longleftrightarrow {\rm R}$ ~~~&
 $\displaystyle{T_B=T_A+{\pi \over 2 }i}$, ~~~~&
  $\displaystyle{R_B=R_A}$,~~~~&
  $R_B^2 - 1 = e^{-\pi i}(1 - R_A^2 )$
\\
\\
~~~~~~
 ${\rm F} \longleftrightarrow {\rm L}$ ~~~&
 $\displaystyle{T_B=-T_L -{\pi \over 2 }i}$,~~~~&
$\displaystyle{R_B=R_L}$,~~~~&
$R_B^2 - 1 = e^{-\pi i}(1 - R_L^2)$,
\\
\\
\hline
\hline
\end{tabular}
\label{tabletwo}
\caption{Continuation of the coordinate variables from F region to R and L regions,
in de~Sitter spacetime.  }
\end{table}
\end{center}

\section{Relationship between the mode functions in the static and future regions} \label{sec:five}
In this section we analytically continue the positive-frequency mode functions $u^F_{p\ell}(R_B,T_B)$ found in the previous section to
the two static regions, the R- and L-regions.  By the observation made in the previous section about the analytic continuation,
i.e.\ $(1 - (-r_R/\eta_R)^2)^{\alpha}\to e^{i\pi\alpha} ((-r_R/\eta_R)^2 - 1)^{\alpha}$ and the formula
$(1-(-r_L/\eta_L)^2)^\alpha \to e^{i\pi\alpha}((-r_L/\eta_L)^2 - 1)^{\alpha}$, which can be derived similarly, we arrive at
the rules stated in Table II.  Using these rules, we find
\begin{eqnarray}
e^{-\frac{\pi |p|}{2}}
U_{|p|\ell}(R_B)e^{-ipT_B} \ \ (\textrm{F-region}) & \leftarrow & \begin{cases} e^{\frac{\pi p}{2}}U^{(S)}_{|p|\ell}(R_A)e^{-ipT_A}\ \
(\textrm{R-region}),\\
e^{-\frac{\pi p}{2}} U^{(S)}_{|p|\ell}(R_L)e^{ipT_L}\ \ (\textrm{L-region}), \end{cases}
\end{eqnarray}
where the functions $U_{|p|\ell}(x)$ for $x>1$ and $U^{(S)}_{|p|\ell}(x)$ for $0\leq x < 1$
are defined by Eqs.~(\ref{U-def-ah}) and (\ref{US-def-ah}), respectively.
Hence, from Eq.~(\ref{def-of-upl}) we readily find
\begin{eqnarray}
u^F_{p\ell}(x_F)  & \leftarrow & \begin{cases}
u_{p\ell}^{(+)}(x) := \dfrac{1}{\sqrt{1- e^{-2\pi|p|}}}\left( v_{p\ell}^{R}(x) + e^{-\pi|p|}\overline{v_{p\ell}^L(x)}\right) & \textrm{for}\ \ p>0,\\
u_{p\ell}^{(-)}(x) := \dfrac{1}{\sqrt{1-e^{-2\pi|p|}}}\left( v_{|p|\ell}^{L}(x) + e^{-\pi|p|}\overline{v_{|p|\ell}^R(x)}\right) & \textrm{for}\ \ p<0,
\end{cases}
\label{Upell-dS}
\end{eqnarray}
where, with $p>0$,
\begin{eqnarray}
v_{p\ell}^R(x) & = & \begin{cases} \sqrt{ 2\sinh (\pi p)} N_{p\ell}U_{|p|\ell}^{(S)}(R_A)e^{-ipT_A} & \textrm{in the R-region},\\
0 & \textrm{in the L-region},\end{cases}\\
v_{p\ell}^L(x) & = & \begin{cases} 0 & \textrm{in the R-region},\\
\sqrt{ 2\sinh (\pi p)} N_{p\ell}U_{|p|\ell}^{(S)}(R_L)e^{-ipT_L} & \textrm{in the L-region}. \end{cases}
\end{eqnarray}
The Klein-Gordon normalization $(v_{p\ell}^R,v_{p'\ell}^R)_{\textrm{KG}} = (v_{p\ell}^{L},v_{p'\ell}^{L})_{\textrm{KG}} = \delta^D(p-p')$ follows from
Eq.~(\ref{Upell-dS}) and $(u^F_{p\ell},u^F_{p'\ell})_{\textrm{KG}} = \delta^D(p-p')$.
The modes $v_{|p|\ell}^R(x)$ coincide with the normalized mode functions
in the right static chart in the literature~\cite{HiguchiA} up to a phase factor as they should.
In exactly the same manner as in the Minkowski case  we find with $p>0$ [see Eqs.~(\ref{defvF1}) and (\ref{defvF2})]
\begin{eqnarray}
v_{p\ell}^R(x)  & \rightarrow & v_{p\ell}^F(x_F) := \frac{1}{\sqrt{1-e^{-2\pi p}}}\left[ u_{p\ell}^F(x_F) - e^{-\pi p}\overline{u_{-p,\ell}^F(x_F)}\right],
\label{vpellR}\\
v_{p\ell}^L(x) & \rightarrow & v_{-p\ell}^F(x_F) : = \frac{1}{\sqrt{1-e^{-2\pi p}}}\left[ u_{-p,\ell}^F(x_F) - e^{-\pi p}\overline{u_{p\ell}^F(x_F)}\right].
\label{vpellL}
\end{eqnarray}
By substituting the definition (\ref{def-of-upl}) 
of $u_p^F(x_F)$ here and comparing the resulting formulas with Eq.~(\ref{UV-reverse}) we find
\begin{eqnarray}
&&  v_{p\ell}^F(x) = \frac{1}{\sqrt{4\pi|p|}}e^{i(\delta'_{|p|}+\delta_{|p|})} V_{|p|\ell}(R_B)e^{-ip T_B},
\end{eqnarray}
where we have defined the phase factor $e^{i\delta'_p}$ by
\begin{eqnarray}
\Gamma(ip) = \sqrt{\frac{\pi}{|p|\sinh |p|}}e^{i\delta'_p}.
\end{eqnarray}

It is clear from the construction of $v_{p\ell}^F(x_F)$ that they are normalized by the radial Klein-Gordon inner product, but it is also easy to verify this fact
directly by examining the behavior of these modes near the horizon.  First we note that by introducing the coordinate $R_{B*}$ as
\begin{eqnarray}
&& R_B = \coth R_{B*},
\end{eqnarray}
the line element of de~Sitter spacetime in the F-region, Eq.~(\ref{F-metric}), becomes
\begin{eqnarray}
ds^2 & = & (R_B^2 -1)\left( - dR_{B*}^2 + dT_B^2\right) + R_B^2 d\Omega_{(2)}^2.
\end{eqnarray}
The radial Klein-Gordon inner product is
\begin{eqnarray}
(v^F_{p\ell},v^F_{p'\ell})_{\textrm{RKG}}
& = &  - i R_{B}^2\int_{-\infty}^\infty dT_B\, \left[ \overline{v^F_{p\ell}(R_B,T_B)}\frac{\partial v^F_{p'\ell}(R_B,T_B)}{\partial R_{B*}}
- \frac{\partial \overline{v_{p\ell}^F(R_B,T_B)}}{\partial R_{B*}} v_{p'\ell}^F(R_B,T_B)\right] \nonumber \\
& = & - {i\over 2|p|}\delta^D(p-p')R_B^2\left[  \overline{V_{|p|\ell}(R_B)}\frac{dV_{|p|\ell} (R_B,T_B)}{dR_{B*}}
- \frac{d\overline{V_{|p|\ell}(R_B)}}{dR_{B*}} V_{|p|\ell}(R_B)\right],
\end{eqnarray}
where we have taken into account the fact that the coordinate $R_{B*}$ decreases towards the future.
This inner product can be evaluated near the horizon, i.e.\ in the limit $R_B \to 1$ (i.e.\ $R_{B*}\to \infty$), by noting that in this limit
\begin{eqnarray}
V_{|p|\ell}(R_B) & \approx & (2R_B)^{-\frac{i|p|}{2}} \approx 2^{-i|p|}e^{i|p|R_{B*}}.  \label{V-behavior}
\end{eqnarray}
Thus we indeed obtain $(v^F_{p\ell}, v^F_{p'\ell})_{\textrm{RKG}} = \delta^D(p-p')$.  We also find from this equation that
\begin{eqnarray}
v^F_{p\ell}(R_B.T_B) & \approx & \frac{1}{\sqrt{4\pi |p|}}e^{i(\delta'_p+\delta_p - |p|\log 2)}e^{i|p|R_{B*} - ipT_B}.
\end{eqnarray}
Thus, the mode functions $v^F_{p\ell}(R_B,T_B)$ are purely left-moving (right-moving) for if $p>0$ ($p<0$).

\section{Entanglement structure of the Bunch-Davies vacuum and alpha-vacua} \label{sec:six}

\begin{figure*}[t]
  \begin{center}
        \includegraphics[width=9cm]{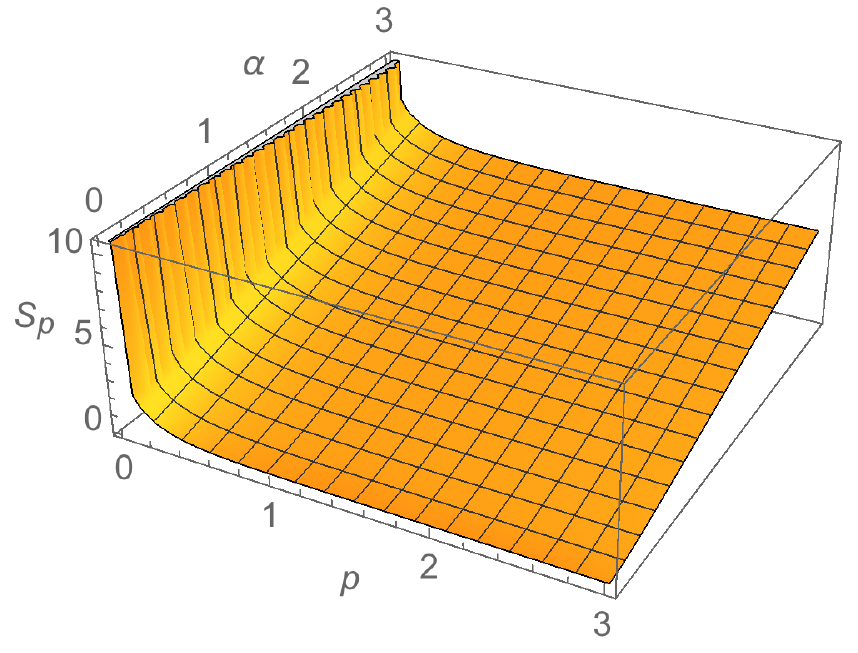}%
  \end{center}
\caption{
  \label{spfigure3} { 3D plot o
    f $s_{p}$ as function of $p$ and $\alpha$ with $\theta=0$.}}
\end{figure*}

Notice the similarity of the relationship between the Minkowski positive-frequency modes and the Rindler modes [see Eq.~(\ref{UF-continueAH})]
 to that between the Bunch-Davies positive-frequency modes and the modes in the static charts [see Eq.~(\ref{Upell-dS})].  This similarity 
makes it straightforward
to write down a relation similar to Eq.~(\ref{M-entangle}) for the Bunch-Davies vacuum.  

We expand the scalar field as
\begin{eqnarray}
\phi(x,\theta,\varphi)  & = & \sum_{\ell, m}\int_0^\infty dp\biggl[ v^R_{p\ell}(x)Y_{\ell m}(\theta,\varphi)\,\hat{a}_{p\ell m}
+ v^L_{p\ell}(x)Y_{\ell m}(\theta,\varphi)\,\hat{a}_{-p,\ell m}  \nonumber \\
&&  ~~~~~~~~~~~~+ \overline{v^R_{p\ell}(x)Y_{\ell m}(\theta,\varphi)}\,\hat{a}_{p\ell m}^\dagger
+ \overline{v^L_{p\ell}(x)Y_{\ell m}(\theta,\varphi)}\,\hat{a}_{-p,\ell m}^\dagger\biggr],
\end{eqnarray}
in the union of the R- and L-regions.
Then, the normalization conditions for the mode functions imply that
$[\hat{a}_{p\ell m},\hat{a}_{p'\ell'm'}^\dagger] = [\hat{b}_{p\ell m},\hat{b}_{p'\ell' m'}^\dagger] = \delta^D(p-p')\delta_{\ell\ell'}\delta_{mm'}$
with all other commutators vanishing.
Recall that the Bunch-Davies vacuum state $|0\rangle_{\textrm{BD}}$ is defined by
the condition that $\hat{b}_{p\ell m}|0\rangle_{\textrm{BD}} = 0$ for all $p$, $\ell$ and $m$, 
where  the operators $\hat{b}_{p\ell m}$ were defined by Eq.~(\ref{F-region-expansion}).
We define the static vacuum state
$|0\rangle_{\textrm{S}}$ by
$\hat{a}_{p\ell m}|0\rangle_{\textrm{S}} = 0$ for all $p$, $\ell$ and $m$.  Then we find
\begin{eqnarray}
|0\rangle_{\textrm{BD}} & = & N \exp\left( \sum_{\ell,m} \int_0^\infty dp\,e^{-\pi p} a_{p\ell m}^{\dagger}a_{-p,\ell m}^{\dagger}\right)
|0\rangle_{\textrm{S}}.
\end{eqnarray}
By discretizing the label $p$, we can write this expression heuristically as
\begin{eqnarray}
|0\rangle_{\textrm{BD}} & \propto & \prod_{p>0,\ell, m}\sum_{n=0}^\infty e^{-\pi p n} |p\ell m, n\rangle_{\textrm{R}}\otimes |p\ell m, n\rangle_{\textrm{L}},
\end{eqnarray}
where $|p\ell m,n\rangle_{\textrm{R}}$   and $|p\ell m,n\rangle_{\textrm{L}}$ are the $n$-particle states for the modes
$v^R_{p\ell}(x_R)Y_{\ell m}(\theta,\varphi)$ in the right static chart and
$v^L_{p\ell}(x_L)Y_{\ell m}(\theta,\varphi)$ in the left static chart, respectively.  Again, it is possible to view this state as a state with entanglement between the
states defined in the right and left static charts.

Next we consider the $\alpha$-vacuum state \cite{Mottola,BA},
which is introduced by adopting the following
mode functions in the F-region as the positive-frequency modes:
\begin{eqnarray}
 \mathcal{U}_{p\ell}^F(x_B)Y_{\ell m}(\theta,\varphi)=\left[ \cosh \alpha\,u^F_{p\ell}(x_B)
  +e^{i\theta}\sinh \alpha\, \overline{u^F_{-p,\ell}(x_B)}\right]Y_{\ell m}(\theta,\varphi).
\label{alpha-def}
\end{eqnarray}
We shall express the $\alpha$-vacuum states $|0\rangle_{\alpha}$ as states with entanglement between the states in the right and left static charts.
By inverting Eq.~(\ref{alpha-def}) and substituting the result into Eqs.~(\ref{vpellR}) and (\ref{vpellL}), we find
\begin{eqnarray}
v_{p\ell}^F(x_B) & = & \alpha_p \mathcal{U}^F_{p\ell}(x_B) + \beta_p \overline{\mathcal{U}^F(x_B)},
\end{eqnarray}
where
\begin{eqnarray}
\alpha_{p}={1\over \sqrt{1-e^{-2\pi |p|}}}  (\cosh\alpha  + e^{-\pi|p|}e^{-i\theta}\sinh\alpha),\nonumber \\
\beta_{p}=  - {1\over \sqrt{1-e^{-2\pi |p|}}}(e^{i\theta}\sinh\alpha +e^{-\pi |p|}  \cosh\alpha).
\end{eqnarray}
Proceeding in the same manner as in the case for the Bunch-Davies vacuum, we find that the $\alpha$-vacuum is expressed in terms of the
static vacuum as follows:
\begin{eqnarray}
  |0\rangle_{\alpha}
  &=& N\exp\left[\int_0^\infty dp ~\gamma_p(\alpha,\theta)
    \hat a_{p\ell m}^\dagger
    \hat a_{-p,\ell m}^\dagger\right]|0\rangle_{(v)}
  \nonumber
  \\
  &\propto &\prod_{{p>0},\ell m}
  \sum_{n=0}^\infty \left[\gamma_p(\alpha,\theta)\right]^{n}
  |p\ell m,n\rangle_{\textrm{S}}\otimes|-p,\ell m,n\rangle_{\textrm{S}},
\end{eqnarray}
where $\gamma_p(\alpha,\theta)=-{\overline{\beta_p}}
/\alpha_p$, $p>0$, is given by
\begin{eqnarray}
  \gamma_p(\alpha,\theta)={e^{-i\theta}\sinh\alpha+e^{-\pi p}\cosh\alpha
    \over
     \cosh\alpha+e^{-\pi p}e^{-i\theta}\sinh\alpha   }.
\label{def-of-gamma}
\end{eqnarray}

The entanglement entropy for a pair of the mode  with given values of $p\,>0$, $\ell$
and $m$ is 
\begin{eqnarray}
  S_{p}=-{\rm Tr_L}[\rho(p)\log \rho(p)]=-\log (1-|\gamma_p|^2)-{|\gamma_p|^2\over 1-|\gamma_p|^2}
\log|\gamma_p|^2,
\label{entropy-result}
\end{eqnarray}
where $\rho(p)$ is the density matrix obtained by tracing out the states with
$p < 0$ or, equivalently, the states in the L-region.
Fig.~\ref{spfigure3} shows the behavior of the entanglement entropy $S_p$
as a function of $p$ and $\alpha$.
We have chosen $\theta=0$, but the behavior is similar for other values of $\theta$
unless $\theta$ takes  values around $\pi$.
This result can be compared with the previous works which computed the
entanglement entropy in de~Sitter spacetime with the two open charts
\cite{Pimentel,KM,Noumi}.
The entanglement entropy for the Bunch-Davies vacuum state,
which is computed with the two static charts, is obtained by the usual
thermal density matrix.
As is clear from Eqs.~(\ref{def-of-gamma}) and (\ref{entropy-result}),
the entanglement entropy for the general $\alpha$-vacuum state
does not depend on the mass of the scalar field, contrary to
these previous works with the  two open charts.
The radius of the spherical surface that divides the spatial slice
into two regions is the Hubble radius for the static charts, while it is
much smaller than the spherical radius for the open charts.
There also exists a region of width of the Hubble radius
between the two open charts \cite{STY}.
It is possible that the difference between our result and
those of Refs.~\cite{Pimentel,KM,Noumi} could be explained by these facts and the fact that
the entanglement entropy is observer-dependent in general.


\section{Summary and Conclusions}
In this paper we investigated the non-interacting massive quantum scalar field in
de~Sitter spacetime, focusing our investigation on the description
of the vacuum state  as an entangled state between the states constructed in
the static charts.
To demonstrate such a description from first principle, we 
constructed positive-frequency modes for the Bunch-Davies
vacuum state in the region to the future
of the two static charts of de~Sitter spacetime.  These positive-frequency
modes have global properties similar to those of positive-frequency modes
in Minkowski spacetime,
which was studied in detail in Ref.~\cite{HIUY}.
The global properties of these modes led directly to the well-known
characterization of the Bunch-Davies
vacuum state as a state with entanglement between the states in the two static charts.
This characterization will be useful for understanding not only the thermal
behavior in vacuum fluctuations but also nonlocal properties
of the quantum field in de~Sitter spacetime.
As an application of this entanglement structure we
computed the entanglement entropy
of a pair of the modes which are entangled in the $\alpha$-vacuum. 
We found that this entropy does not depend on the mass of the field, contrary to the results
with the two open charts in the literature.

The description of the Minkowski vacuum state
as an entangled state between the two Rindler wedges is known to be
useful for understanding
the quantum radiation
produced by a uniformly accelerating detector coupled to vacuum fluctuation
of a field in Minkowski spacetime~\cite{HIUY},
which will be important for testing the Unruh effect.
A similar quantum radiation has been discovered in the model consisting of
a uniformly accelerating detector coupled to vacuum fluctuation
of a field in de~Sitter spacetime \cite{Yamaguchi1}. It has been shown that there exists
a similar theoretical structure in the formulas for the quantum radiation
in Minkowski spacetime and de~Sitter spacetime.
The results in the present paper will be useful for understanding the origin of the
quantum radiation in a similar context.

\vspace{2mm}

\noindent
{\it Acknowledgments.}---
This work was supported by MEXT/JSPS KAKENHI Grant Numbers 15H05895,
17K05444, and 17H06359 (KY). We thank K.~Ueda, Y.~Nambu, J.~Soda,
S.~Kanno, S.~Iso, S-Y.~Lin, T.~Markkanen,
and M.~Sasaki for useful discussions
related to the topic in the present paper.
K.~Y.\ is grateful for the warm hospitality of the Mathematics Department
of the University of York, where part of this work was carried out.


\end{document}